\documentclass[11pt,a4paper]{article}
\usepackage[hyperref]{acl2020}
\usepackage{times}
\usepackage{latexsym}

\usepackage{times}  %Required
\usepackage{helvet}  %Required
\usepackage{courier}  %Required
\usepackage{url}  %Required
\usepackage{graphicx}  %Required

\usepackage{amssymb}

\usepackage{enumitem}
\usepackage{subfig} 
\usepackage{amsmath}
\usepackage{booktabs}

\usepackage{algorithmicx}
\usepackage{algpseudocode}

\usepackage{algorithm}
\usepackage{multirow} 
\usepackage{amsmath} 
\usepackage{xcolor}

\usepackage{microtype}

\aclfinalcopy % Uncomment this line for the final submission
\title{Bipartite Flat-Graph Network for Nested Named Entity Recognition}

\author{Ying Luo \and Hai Zhao\thanks{$\ $ Corresponding author. This paper was partially supported by National Key Research and Development Program of China (No. 2017YFB0304100), Key Projects of National Natural Science Foundation of China (U1836222 and 61733011).}\\
Department of Computer Science and Engineering, Shanghai Jiao Tong University \\
Key Laboratory of Shanghai Education Commission for Intelligent Interaction \\ and Cognitive Engineering, Shanghai Jiao Tong University, Shanghai, China\\
MoE Key Lab of Artificial Intelligence, AI Institute, Shanghai Jiao Tong University, Shanghai, China\\
{\tt kingln@sjtu.edu.cn, zhaohai@cs.sjtu.edu.cn}
}

\date{}

\begin{document}
\maketitle
\begin{abstract}

In this paper, we propose a novel bipartite flat-graph network (BiFlaG) for nested named entity recognition (NER), which contains two subgraph modules: a flat NER module for outermost entities and a graph module for all the entities located in inner layers.
Bidirectional LSTM (BiLSTM) and graph convolutional network (GCN) are adopted to
jointly learn flat entities and their inner dependencies.
Different from previous models, which only consider the unidirectional delivery of information from innermost layers to outer ones (or outside-to-inside), our model effectively captures the bidirectional interaction between them. 
We first use the entities recognized by the flat NER module to construct an entity graph, which is fed to the next graph module. 
The richer representation learned from graph module carries the dependencies of inner entities and can be exploited to improve outermost entity predictions.
Experimental results on three standard nested NER datasets demonstrate that our BiFlaG outperforms previous state-of-the-art models.

\end{abstract}

\section{Introduction}

\begin{figure}[!t]
  \centering 
  \includegraphics[scale=0.58]{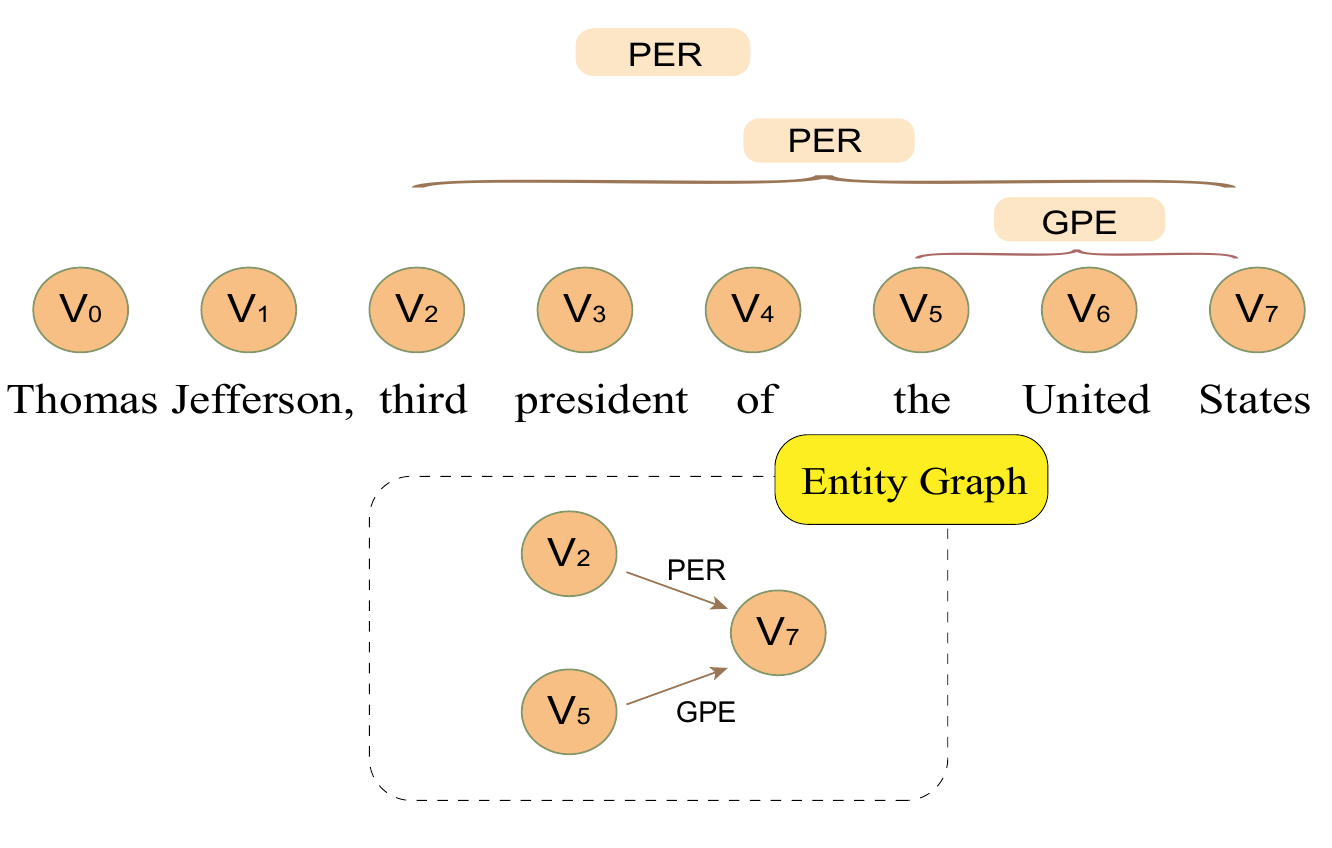}
    \caption{An example of nested named entity mentions. Solid lines connect the starting and ending indices of inner nested entities.}\label{example}
\end{figure} 

Named entity recognition (NER) aims to identify words or phrases that contain the names of pre-defined categories like location, organization or medical codes. Nested NER further deals with entities that can be nested with each other, such as \textit{the United States} and \textit{third president of the United States} shown in Figure \ref{example}, such phenomenon is quite common in natural language processing (NLP). 

NER is commonly regarded as a sequence labeling task \cite{lample2016neural,ma2016end,peters2017semi}. These approaches only work for non-nested entities (or flat entities), but neglect nested entities. There have been efforts to deal with the nested structure.
\citeauthor{ju2018neural} \citeyear{ju2018neural} introduced a layered sequence labeling model to first recognize innermost entities, and then feed them into the next layer to extract outer entities. However, this model suffers from obvious error propagation. The wrong entities extracted by the previous layer will affect the performance of the next layer. Also, such layered model suffers from the sparsity of entities at high levels. For instance, in the well-known ACE2005 training dataset, there are only two entities in the sixth level.
\citeauthor{sohrab2018deep} \citeyear{sohrab2018deep} proposed a region-based method that enumerates
all possible regions and classifies their entity types. However, this model may ignore explicit boundary information.
\citeauthor{zheng2019boundary} \citeyear{zheng2019boundary}  combined the layered sequence labeling model and region-based method to locate the entity boundary first, and then utilized the region classification model to predict entities. This model, however, cares less interaction among entities located in outer and inner layers. 

In this paper, we propose a bipartite flat-graph network (BiFlaG) for nested NER, which models a nested structure containing arbitrary many layers into two parts: outermost entities and inner entities in all remaining layers.
For example, as shown in Figure \ref{example}, the outermost entity \textit{Thomas Jefferson, third president of the United States} is considered as a flat (non-nested) entity, while \textit{third president of the United States} (in the second layer)  and \textit{the United States} (in the third layer) are taken as inner entities. 
The outermost entities with the maximum coverage are usually identified in the flat NER module, which commonly adopts a sequence labeling model. 
All the inner entities are extracted through the graph module, which iteratively propagates information between the start and end nodes of a span using graph convolutional network (GCN) \cite{kipf2016semi}.
The benefits of our model are twofold:
(1) Different from layered models such as \cite{ju2018neural}, which suffers from the constraints of one-way propagation of information from lower to higher layers, our model fully captures the interaction between outermost and inner layers in a bidirectional way. Entities extracted from the flat module are used to construct entity graph for the graph module.
Then, new representations learned from graph module are fed back to the flat module to  improve  outermost entity predictions.
Also, merging all the entities located in inner layers into a graph module can effectively alleviate the sparsity of entities in high levels.
(2) Compared with region-based models \cite{sohrab2018deep,zheng2019boundary}, our model makes full use of the sequence information of outermost entities, which take a large proportion in the corpus.

The main contributions of this paper can be summarized as follows:
\begin{itemize}
    \item We introduce a novel bipartite flat-graph network named BiFlaG for nested NER, which incorporates a flat module for outermost entities and a graph module for inner entities.
    \item Our BiFlaG fully utilizes the sequence information of outermost entities and meanwhile bidirectionally considers the interaction between outermost and inner layers, other than unidirectional delivery of information.
    \item With extensive experiments on three benchmark datasets (ACE2005, GENIA, and KBP2017), our model outperforms previous state-of-the-art models under the same settings.
\end{itemize} 

\section{Model}

\begin{figure*}[!t]
  \centering 
  \includegraphics[scale=0.68]{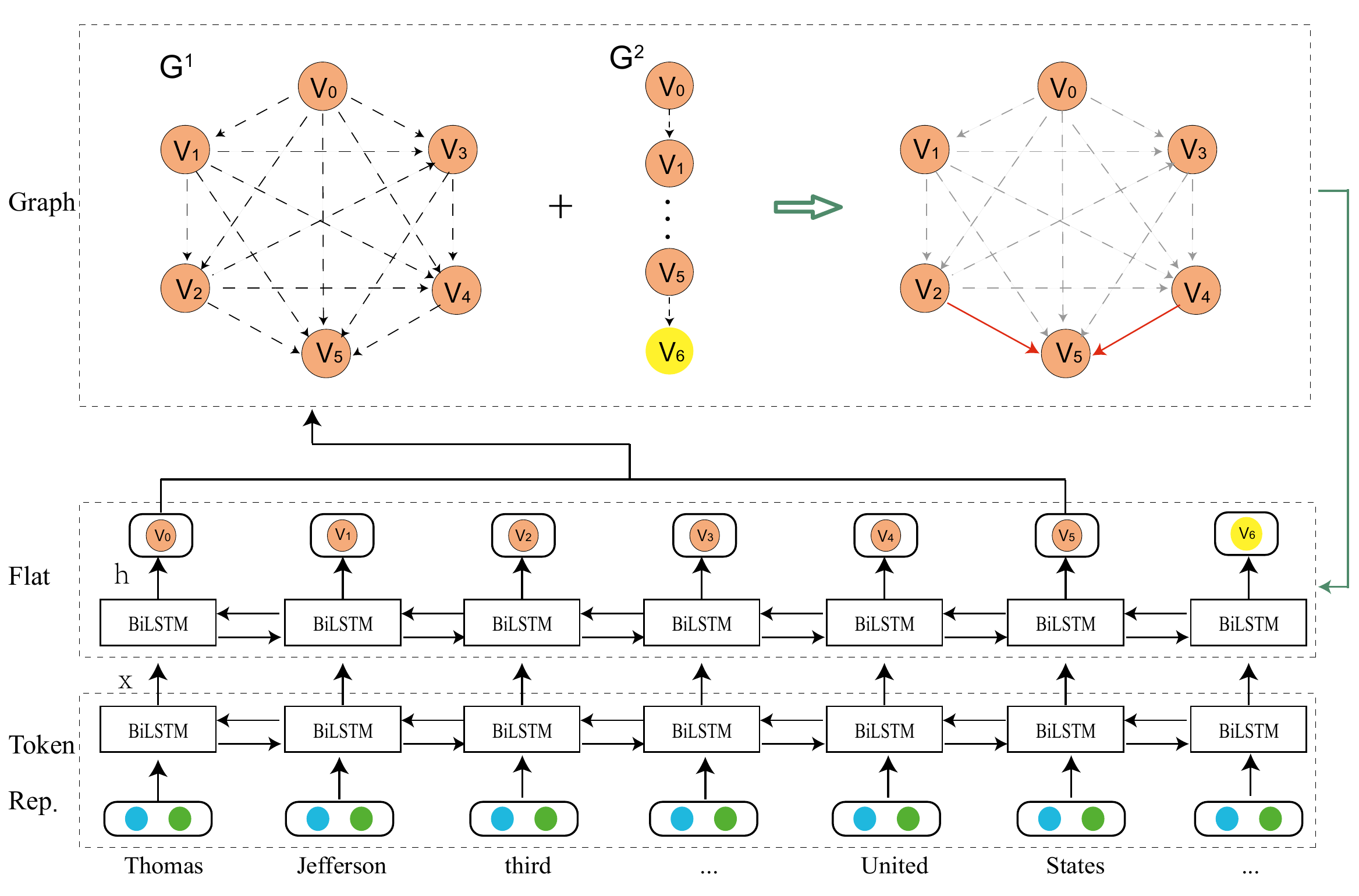} 
    \caption{The framework of our BiFlaG model. $G^1$ and $G^2$ are entity graph and adjacent graph created for GCN, each dashed line connects the start and end nodes for a potential entity. Solid red lines indicate inner entities recognized by the graph module.}\label{Overview}
\end{figure*}

Our BiFlaG includes two subgraph modules, a flat NER module and a graph module to learn outermost and inner entities, respectively. Figure \ref{Overview} illustrates the overview of our model.
For the flat module, we adopt BiLSTM-CRF to extract flat (outermost) entities, and use them to construct the entity graph $G^1$ as in Figure \ref{Overview}.
For the graph module, we use GCN which iteratively propagates information between the start and end nodes of potential entities to learn inner entities. Finally, the learned representation from the graph module is further fed back to the flat module for better outermost predictions.

\subsection{Token Representation}
Given a sequence consisting of $N$ tokens $\{t_1, t_2, ..., t_N\}$, for each token $t_i$, we first concatenate the word-level and character-level embedding  $t_i = [w_i; c_i$], $w_i$ is the pre-trained word embedding, character embedding $c_i$ is learned following the work of \cite{xin2018learning}. Then we use a BiLSTM to capture sequential information for each token $x_i = \Call{BiLSTM}{t_i}$.
We take $x_i$ as the word representation and feed it to subsequent modules. 

\subsection{Flat NER Module}
We adopt BiLSTM-CRF architecture \cite{lample2016neural, ma2016end, yang2018ncrf,luo2019hierarchical} in our flat module to recognize flat entities, which consists of a bidirectional LSTM (BiLSTM) encoder and a conditional random field (CRF) decoder.

\textbf{BiLSTM} 
% \cite{hochreiter1997long} introduce a memory-cell to  capture long-range dependencies.
captures bidirectional contextual information of sequences and can effectively represent the hidden states of words in context.
 BiLSTM represents the sequential information at each step, the hidden state $h$ of BiLSTM can be expressed as follows.
\begin{equation}
    \begin{aligned}
        \overrightarrow{h_i} &= LSTM(x_i, \overrightarrow{h}_{i-1}; \overrightarrow{\theta}) \\
        \overleftarrow{h_i} &= LSTM(x_i, \overleftarrow{h}_{i-1}; \overleftarrow{\theta}) \\
        h_i &= [\overrightarrow{h_i}; \overleftarrow{h_i}]
    \end{aligned}
\end{equation}
where $\overrightarrow{\theta}$ and $\overleftarrow{\theta}$ are trainable parameters. $\overrightarrow{h_i}$ and $\overleftarrow{h_i}$ respectively denote the forward and backward  context representations of token $t_i$.
The output of BiLSTM $H= \{h_1, h_2, ..., h_N\}$ is further fed into the CRF layer.

\textbf{CRF}
 \cite{Lafferty:01} has been widely used in state-of-the-art NER models \cite{lample2016neural,ma2016end,yang2018ncrf} to help make better decisions,  which considers strong label dependencies by adding transition scores between neighboring labels.
Viterbi algorithm is applied to search for the label sequence with  highest probability during the decoding process.
For $y=\{y_1,...,y_N\}$ being a sequence of predictions with length $N$. Its score is defined as follows.
\begin{equation}  
s(x, y) = \sum_{i=0}^{N-1} T_{y_i, y_{i+1}} + \sum_{i=1}^{N} P_{i, y_i}
\end{equation}
where $T_{y_i, y_{i+1}}$ represents the transmission score  from  $y_i$ to $y_{i+1}$,
$ P_{i, y_i}$  is the score of the $j^{th}$ tag of the $i^{th}$ word from BiLSTM encoder.

CRF model defines a family of conditional probability $p(y|x)$ over all possible tag sequences $y$:
% A softmax over all possible tag sequences in the sentences generates a probability for the sequence $y$:
\begin{equation} 
p(y|x) = \frac{\exp^{s(x,y)}}{\sum_{\tilde{y}\in y}\exp^{s(x, \tilde{y})}} 
\end{equation}
during training phase, we consider the maximum log probability
of the correct predictions. 
While decoding, we search the tag sequences with maximum score:
\begin{equation}  
y^{*} = \arg\max_{\tilde{y} \in y} score(x, \tilde{y})
\end{equation}

\subsection{Graph Module}
Since the original input sentences are plain texts without inherent graphical structure, we first construct graphs based on the sequential information of texts and the entity information from the flat module.
Then, we apply GCN \cite{kipf2016semi, qian2018graphie} which propagates information between neighboring nodes in the graphs, to extract the inner entities.

\textbf{Graph Construction.}
We create two types of graphs for each sentence as in Figure \ref{Overview}. Each graph is defined as $G=(V, E)$, where $V$ is the set of nodes (words), $E$ is the set of edges.
\begin{itemize}
    \item Entity graph $G^1$: for all the nodes in an extracted entity extracted from the flat module, edges are added between any two nodes $e_{ij} = (v_i, v_j)$, where $ start \le i < j \le end$, as shown in Figure \ref{Overview}, allowing the outermost entity information to be utilized.
    \item Adjacent graph $G^2$: for each pair of adjacent words in the sentence, we add one directed edge from the left word to the right one, allowing local contextual information to be utilized.
\end{itemize}

\textbf{Bi-GCN.}
In order to consider both incoming and outgoing features for each node, we follow the work of \cite{marcheggiani2017encoding, fu2019graphrel}, which uses Bi-GCN to extract graph features. 
Given a graph $G = (V, E)$, and the word representation $X = \{x_1, x_2,..., x_N\}$, the graph feature $f \in \mathbb{R}^{N \times d_f}$ learned from Bi-GCN is expressed as follows.

\begin{equation}
    \begin{aligned}
        \overrightarrow{f_i} &= ReLU(\sum_{{e_{ij}} \in E }(\overrightarrow{W_f}x_j+ \overrightarrow{b_f})) \\
        \overleftarrow{f_i} &= ReLU(\sum_{{e_{ji}} \in E }(\overleftarrow{W_f}x_j+ \overleftarrow{b_f})) \\ 
        f_i &= [\overrightarrow{f_i}; \overleftarrow{f_i}]
    \end{aligned}
\end{equation}
where $W_f \in \mathbb{R}^{d_x \times d_f }$ and $b_f \in \mathbb{R}^{d_f}$ are trainable parameters, $d_x$ represents the dimension of word representation, $d_f$ is the hidden size of GCN, $ReLU$ is the non-linear activation function. ${e_{ij}}$ represents the edge outgoing from token $t_i$, and ${e_{ji}}$ represents the edge incoming to token $t_i$. 

The features of the two graphs are aggregated to get impacts of both graphs
\begin{equation}
    f = W_c(f^1 \oplus f^2) + b_c
\end{equation}
where $W_c \in \mathbb{R}^{2d_f \times d_f}$ is the weight to be learned, $b_c \in \mathbb{R}^{d_f}$ is a bias parameter. $f^1$ and $f^2$ are graph features of $G^1$ and $G^2$, respectively.

After getting the graph representation $F = \{f_1, f_2, ..., f_N\}$ from Bi-GCN, we learn the entity score $M \in \mathbb{R}^{N \times N \times L}$ for inner layers as
\begin{equation}
    \begin{aligned}
        M_{ij} = softmax(W_3ReLU(W_1f_i \oplus W_2f_j)) \\
    \end{aligned}
\end{equation}
where $W_{1}, W_2 \in \mathbb{R}^{d_f \times {d_f}/2}$, $W_{3}\in \mathbb{R}^{d_f \times L}$, $L$ is the number of entity types. $M_{ij} \in \mathbb{R}^L$ represents the type probability for a span starts from token $t_i$ and ends at token $t_j$.

%  Cross Entropy is utilized for GCN training over inner entities.
 For inner entities, we define the ground truth entity of word pair $(t_i, t_j)$ as $\hat{M}_{ij}$, where $t_i$ and $t_j$ are start and end nodes of a span.
Cross Entropy (CE) is used to calculate the loss
\begin{equation}
    \begin{aligned}
        L_{inner} = & - ( \sum(\hat{M}_{ij}\text{log}(M_{ij})) \cdot I(\rm O) + \\ 
        & \lambda_1 \cdot \sum(\hat{M}_{ij}\text{log}(M_{ij})) \cdot (1 - I(\rm O)))
    \end{aligned}
\end{equation}
where $M_{ij} \in \mathbb{R}^{L}$ denotes the entity score in the graph module. $I(\rm O)$ is a switching function to distinguish the loss of non-entity 'O' and other entity types. It is defined as follows.
\begin{equation}  
I(\rm O) =\left\{
\begin{aligned}
1, \text{if type $=$ '\rm O'} \\
0, \text{if type $\neq$ '\rm O'}
\end{aligned}
\right.  
\end{equation} 
$\lambda_1$ is the bias weight. The larger $\lambda_1$ is, the greater impacts of entity types, and the smaller influences of non-entity 'O' on the graph module.

\renewcommand{\algorithmicrequire}{\textbf{Input:}}
\renewcommand{\algorithmicensure}{\textbf{Output:}}

 \begin{algorithm}[t!]
    \caption{Bipartite Flat-Graph Algorithm}
    \begin{algorithmic}[1] 
        \Require {word representations $X =\{x_1,..,x_N\}$, 
        \newline number of entity types $L$ \newline the dimension of word embeddings $d_x$,
        \newline the hidden size of GCN $d_f$}
        \Ensure all the entities in this sequence 
        \For {numbers of training iterations}
            \State $y \gets $ \Call{BiLSTM-CRF}{$X$}
            \State create entity graph $G^1$ based on $y$ 
            \State $F_{N \times d_f} \gets $ \Call{Bi-GCN}{$X, G^1$}
            \State $M_{N \times N \times L} \gets$ \Call{Linear}{$F \times F $}
            % \State $R_{N \times N} \gets $ \Call{max}{M} 
            \State transform $M$ to graph $G^3$ by Eq.(10)
            \State $X^{new} \gets $ \Call{Bi-GCN}{$X, G^3$}
            \State $y^{new} \gets $ \Call{BiLSTM-CRF}{$X^{new}$}
            \State entity set $T$ $\gets$ entities in $M$ and $y^{new}$
        \EndFor
        \State \Return entity set $T$
    \end{algorithmic}
    \label{algorithm}
\end{algorithm}

\begin{table*}[!h] 
\resizebox{1.0\textwidth}{!}{
\centering
\begin{tabular}{r|rr|rr|rr|rr|rr|rr}
\toprule
& \multicolumn{6}{c|} {\textbf{ACE2005}} & \multicolumn{6}{c}{\textbf{GENIA}}\\
 & Train  &(\%) & Dev & (\%) & Test & (\%)   & Train & (\%) & Dev & (\%) & Test & (\%) \\
\hline   
\# sentences & 7,285 && 968 && 1,058 &&  15,022 & & 1,669 && 1,854& \\
with o.l. & 2,820 &(39) & 356 & (37)& 344 & (33) &3,432&(23) & 384 & (23) & 467 & (25) \\
\hline  
\# mentions & 24,827 & & 3,234 &  &3,028 && 47,027 & & 4,469 && 5,596 &\\ 
outermost entity & 18,656 & (75) & 2,501 & (77) & 2,313 & (76) &42,558&(90)&4,030& (90) &4,958 &(89) \\ 
inner entity & 6,171 & (25) & 733 & (23) & 715 & (24) & 4,469& (10)&439&(10) &642&(11)\\
\bottomrule 
\end{tabular}}
\caption{Statistics of the datasets used in our experiments: ACE2005 and KBP2017. o.l.: overlapping mentions.} 
\label{datset1}
\end{table*} 

\subsection{BiFlaG Training}
The entity score $M$ in Eq.(7) carries the type probability of each word pair in the sentence. To further consider the information propagation from inner entities to outer ones, we use Bi-GCN to generate new representations from entity score $M$ for the flat module. The largest type score $r_{ij}$ of the word pair $(t_i, t_j)$ indicates  whether this span is an entity or non-entity and the confidence score of being such type, which  is obtained by a max-pooling operation:

\begin{equation}
r_{ij} =\left\{
\begin{aligned}
max(m_{ij}), \text{if type $\neq$ 'O'} \\
0, \text{if type $=$ 'O'}
\end{aligned}
\right. 
\end{equation}
where \textit{type} represents the entity type or non-entity 'O' corresponding to the maximum type score. 
When the corresponding type is \textit{O}, there exits no dependencies between $t_i$ and $t_j$, thus we set $r_{ij}$ to 0.
A new graph that carries the boundary information of inner entities is defined as $G^3 = (V, E)$, where $r_{ij} \in E$.

 The new representation used to update flat module consists of two parts. The first part carries the previous representation of each token
\begin{equation}
    \alpha_i^1 = W_rx_i + b_r
\end{equation}
where $W_r \in \mathbb{R}^{d_x \times d_f}$, $b_r \in \mathbb{R}^{d_f}$.
The second part aggregates inner entity dependencies of the new graph $G^3$
\begin{equation}
    \alpha_i^2 = \Call{Bi-GCN}{x_i, G^3}
\end{equation}
Finally, $\alpha_i^1$ and $\alpha_i^2$ are added to obtain the new representation
\begin{equation}
    x_i^{new} = \alpha_i^1 +  \alpha_i^2
\end{equation}
 $x_i^{new}$ is fed into the flat module to update the parameters and extract better outermost entities.
 
 For outermost entities, we use the BIOES sequence labeling scheme and adopt CRF to calculate the loss. The losses corresponding to the two representations ($X$ and $X^{new}$) are added together as the outermost loss
\begin{equation}
    L_{outer} = CRF_{X} + CRF_{X^{new}}
\end{equation}
 
Entities in the sequence are divided into two disjoint sets of outermost and inner entities, which are modeled by flat module and graph module, respectively.
Entities in each module share the same neural network structure.
Between two modules, each entity in the flat module is either an independent node, or interacting with one or more entities in the graph module. Therefore, Our BiFlaG is indeed a bipartite graph.
Our complete training procedure for BiFlaG is shown in Algorithm \ref{algorithm}.

\subsection{Loss Function}
Our BiFlaG model predicts both outermost and inner entities. 
The total loss is defined as
\begin{equation}
    L =  L_{outer} + \lambda_2 L_{inner}
\end{equation}
where $\lambda_2$ is a weight between loss of flat module and graph module. We minimize this total loss during training phase.

\begin{table*}[!t] 
\resizebox{1.0\textwidth}{!}{
\centering
\begin{tabular}{l||ccc||ccc||ccc}
\toprule
 & \multicolumn{3}{c||} {\textbf{ACE2005}} & \multicolumn{3}{c||} {\textbf{GENIA}}  & \multicolumn{3}{c}{\textbf{KBP2017}}\\
 \cline{2-10}
Model & {P} & {R} & {F1}   & {P} & {R} & {F1}  & {P} & {R} & {F1}  \\
\hline   
LSTM-CRF \cite{lample2016neural} & 70.3 & 55.7& 62.2 &75.2& 64.6& 69.5
&71.5 &53.3& 61.1 \\
Multi-CRF & 69.7 & 61.3 & 65.2 & 73.1 & 64.9 &68.8& 69.7 &60.8 &64.9 \\
\hline
layered-CRF \cite{ju2018neural} & 74.2 & 70.3 & 72.2 & 78.5 & 71.3 &74.7 & -&-&-\\ 
\hline
LSTM. hyp \cite{katiyar2018nested}&70.6 &70.4 &70.5 &79.8 &68.2 &73.6& -&-&- \\ 
Segm. hyp [\textbf{POS}] \cite{wang2018neural} & 76.8 & 72.3 & 74.5 & 77.0 & 73.3& 75.1$^*$& 79.2& 66.5 &72.3 \\
\hline
Exhaustive \cite{sohrab2018deep} \footnotemark[4] & -&-&- & 73.3 & 68.3 & 70.7 & -&-&-  \\ 
Anchor-Region [\textbf{POS}] \cite{lin2019sequence} & 76.2 & 73.6 &74.9 &75.8 &73.9& 74.8  &77.7 &71.8& 74.6$^*$ \\
Merge \& Label \cite{fisher2019merge} & 75.1 &74.1 &74.6$^{\dagger}$ & -&-&-& -&-&-\\
Boundary-aware \cite{zheng2019boundary} & -&-&- & 75.9 & 73.6 & 74.7$^{\dagger}$ &-&-&-\\
GEANN [\textbf{Gazetter}] \cite{lin2019gazetteer} & 77.1 & 73.3 & \textbf{75.2$^*$} &-&-&- &-&-&- \\
\hline 
KBP2017 Overview \cite{ji2017overview} & - &- &- &- &- &- &72.6 &73.0 &72.8$^{\dagger}$\\
\hline\hline
BiFlaG & 75.0 & 75.2 & {75.1} &  77.4& 74.6 & \textbf{76.0} & 77.1 & 74.3 & \textbf{75.6} \\ 
& &&(-0.1$^*$)&&&(+0.9$^*$) &&&(+1.0$^*$)\\
\bottomrule
\end{tabular}}
\caption{Experimental results\footnotemark[5] on ACE2005, GENIA and KBP2017 datasets. POS and Gazetteer indicates using additional POS tags and gazetteers.
$^{\dagger}$ represents previous state-of-the-art results under the same settings with our experiments, $^*$ represents state-of-the-art results with POS tags or gazetteers, values in parentheses are also compared with them. } 
\label{result}
\end{table*} 

\section{Experiment}

\subsection{Dataset and Metric}
We evaluate our BiFlaG on three standard nested NER datasets: GENIA, ACE2005, and TACKBP2017 (KBP2017) datasets, which  contain 22\%, 10\% and 19\% nested mentions, respectively. Table \ref{datset1} lists the concerned data statistics.

\textbf{GENIA} dataset \cite{kim2003genia} is based on the GENIAcorpus3.02p\footnote{http://www.geniaproject.org/genia-corpus/pos-annotation}. We use the same setup as previous works \cite{finkel2009nested, lu2015joint, lin2019sequence}. This dataset contains 5 entity categories
and is split into 8.1:0.9:1 for training, development and test.

\textbf{ACE2005}\footnote{https://catalog.ldc.upenn.edu/LDC2006T06 (ACE2005)} \cite{walker2006ace} contains 7 fine-grained entity categories. We preprocess the dataset following
the same settings of \cite{lu2015joint,wang2018neural, katiyar2018nested, lin2019sequence} by keeping files from bn, nw
and wl, and splitting these files into training, development and
test sets by 8:1:1, respectively. 

\textbf{KBP2017} Following \cite{lin2019sequence}, we evaluate our model on the 2017 English evaluation dataset (LDC2017E55). The training and development sets contain previous RichERE annotated datasets (LDC2015E29,
LDC2015E68, LDC2016E31 and LDC2017E02).
The datasets are split into 866/20/167 documents
for training, development and test, respectively.

\textbf{Metric} Precision ($P$), recall ($R$) and F-score ($F_1$) are
used to evaluate the predicted entities. 
An entity is confirmed correct if it exists in the
target labels, regardless of the layer at which the model makes this prediction.

\subsection{Parameter Settings} 
Our model \footnote{Code is available at: https://github.com/cslydia/BiFlaG.} is based on the framework of \cite{yang2018ncrf}.
We conduct optimization with the stochastic gradient descent (SGD) and Adam for flat  and GCN modules, respectively. For GENIA dataset, we use the same 200-dimension pre-trained word embedding as \cite{ju2018neural, sohrab2018deep, zheng2019boundary}. For ACE2005 and KBP2017 datasets,
we use the publicly available pre-trained 100-dimension GloVe \cite{pennington2014glove} embedding.
We train the character embedding as in \cite{xin2018learning}. 
The learning rate is set to 0.015 and 0.001 for flat and GCN modules, respectively. We apply
dropout to embeddings and the hidden states  with a rate of 0.5.
The hidden sizes of BiLSTM and GCN are both set to 256. The bias weights $\lambda_1$ and $\lambda_2$ are both set to 1.5.

\footnotetext[4]{This result is reported by \cite{zheng2019boundary}, consistent with our own re-implemented results.}

% \footnotetext[5]{We do not conduct experiments on pre-trained language model
% BERT \cite{devlin2018bert}, since it has a certain number of layers, the combination of different layers has a great impact on NER results, and eventually leads to diverse and impalpable results. For example, the results of \cite{lin2019gazetteer} and \cite{fisher2019merge} are 75.2/80.1, 74.7/82.4 $F_1$ without/with BERT. In this work, we focus on how to strengthen the model itself rather than about introducing external resources, which lets us exclude the BERT-related evaluation.}

\subsection{Results and Comparisons} 
Table \ref{result} compares our model to some existing state-of-the-art
approaches on the three benchmark datasets. Given only standard training
data and publicly available word embeddings, the results in Table 2 show that our model outperforms all these  models. Current state-of-the-art results on these datasets are tagged with $^\dagger$ in Table \ref{result}, we make improvements of 0.5/1.3/2.8 $F_1$ on ACE2005, GENIA, and KBP2017 respectively. 
KBP2017 contains much more entities than ACE2005 and GENIA. The number of entities on  test set is four times  that of ACE2005. Our model has the most significant improvement on such dataset, proving the effectiveness of our BiFlaG model.
More notably, our model without POS tags surpasses the previous models \cite{wang2018neural, lin2019sequence}, which use POS tags as additional representations on all three datasets.
Besides, \cite{lin2019gazetteer} incorporate gazetteer information on ACE2005 dataset, our model also makes comparable results with theirs.
 Other works like \cite{strakova2019neural} \footnote{Their reported results are 75.36 and 76.44 trained on concatenated train+dev sets on ACE2005 and GENIA, respectively. They also use lemmas and POS tags as additional features.}, which train their model on both training and development sets, are thus not comparable to our model directly.

\begin{table*}[h!]
%\resizebox{1.0\textwidth}{!}{
\centering
\begin{tabular}{c ccccc ccc cccc cr}
\hline
& \multicolumn{3}{c} {{Our model}} && \multicolumn{3}{c} {Boundary-aware } && \multicolumn{3}{c} {{Layered-CRF}}  \\ 
\cline{2-4}\cline{6-8}\cline{10-12}
Category & P & R &F && P & R &F && P & R &F  &&Num.\\
\hline
DNA &  72.7 & 72.7  & \textbf{72.7} && 73.6 &67.8 &70.6 && 74.4 &69.7& 72.0 && 1,290 \\ 
RNA & 84.4 & 84.4 & {84.4} && 82.2& 80.7& 81.5 && 90.3 &79.5 &\textbf{84.5} && 117\\ 
Protein & 79.5 & 76.5 & \textbf{78.0}  && 76.7& 76.0 &76.4 && 80.5& 73.2& 76.7 && 3,108 \\ 
Cell Line & 75.9 & 67.6 & \textbf{71.5} && 77.8 &65.8 &71.3 && 77.8& 65.7& 71.2 && 462 \\ 
Cell Type & 76.7 & 72.4  &\textbf{74.4} && 73.9& 71.2 &72.5&& 76.4& 68.1& 72.0 && 619 \\
\hline 
Overall & 77.4 & 74.6 & \textbf{76.0} && 75.8 & 73.6 &74.7 && 78.5& 71.3 &74.7 && 5,596\\
\bottomrule
\end{tabular}
\caption{Our results on five categories compared to \cite{zheng2019boundary} and \cite{ju2018neural} on GENIA dataset.}
\label{genia}
\end{table*}  

Table \ref{genia} makes a detailed comparison on the five categories of GENIA test dataset with a layered model \cite{ju2018neural} and a region-based model \cite{zheng2019boundary}.
Compared with region-based model, layered model seems to have higher precision and lower recall, for they are subject to error propagate, the outer entities will not be identified if the inner ones are missed. Meanwhile, region-based model suffers from low precision, as they may generate a lot of candidate spans. By contrast, our BiFlaG model well coordinates precision and recall. The entity types \textit{Protein} and \textit{DNA} have the most nested entities on GENIA dataset, the improvement of our BiFlaG on these two entity types is remarkable, which can be attributed to the interaction of nested information between the two subgraph modules of our BiFlaG.

\subsection{Analysis of Each Module}
Table \ref{layer} evaluates the performance of each module on ACE2005 and GENIA datasets. Our flat module performs well on both datasets for outermost entity recognition. However, the recall of the inner entities is low on GENIA dataset.
According to the statistics in Table \ref{datset1}, only 11\% of the entities on GENIA are located in inner layers, while on ACE2005 dataset, the proportion is 24\%.
It can be inferred that the sparsity of the entity distribution in inner layers has a great impact on the results. If these inner entities are identified at each layer, the sparsity may be even worse.
We can enhance the impact of sparse entities by increasing the weight $\lambda_1$ in Eq.(14), but this may hurt precision, we set $ \lambda_1=1.5 $ to have a better tradeoff between precision and recall.

\begin{table}[h!]
\resizebox{1.01\columnwidth}{!}{
\centering
\begin{tabular}{c| ccc| ccc}
\toprule
& \multicolumn{3}{c|} {{ACE2005}} & \multicolumn{3}{c} {GENIA } \\ 
& P & R & F & P & R & F \\
\hline
Outermost & 73.7 &75.0&74.3 & 78.4 & 78.9  & 78.7 \\ 
Inner & 58.3 & 55.2  & 56.7 &50.9 &34.7&41.2 \\
% \hline
% Overall& 75.0&75.2&75.1 & 77.4 & 74.6 & 76.0\\
\bottomrule
\end{tabular}}
\caption{Performance of each module on ACE2005 and GENIA datasets.}
\label{layer}
\end{table}

 \begin{table}[!t] 
\centering
\resizebox{1.03\columnwidth}{!}{
\begin{tabular}{rccc}
\toprule
\textbf{\textbf{}} & ACE2005 & GENIA & KBP2017\\
\midrule 
\textbf{Flat $\rightarrow$  Grpah }  \\ 
    no graph & 73.4 & 74.4 & 74.0 \\ 
    adjacent graph & 73.8 & 74.9 & 74.7\\ 
    entity graph & 74.8 & 75.5 & 75.2\\ 
\hline
    both graphs & 75.1 & 76.0 & 75.6\\ 
\midrule
\textbf{Graph $\rightarrow$  Flat} \\ 
    without & 74.3 & 74.5 & 75.1\\
\hline
    with & 75.1 & 76.0 & 75.6\\
\bottomrule
\end{tabular}}
\caption{Ablation study on the three benchmark datasets.}
\label{ablation}
\end{table}

\subsection{Analysis of Entity Length}
We conduct additional experiments on ACE2005 dataset to detect the effect of the lengths of the outermost entities on the extraction of their inner entities as shown in Table \ref{lengthwise}.
Our flat module can well predict outermost entities which account for a large proportion among all types of entities.
In general, the performance of inner entities is affected by  the extracting performance and length of their outermost entities.
A shorter outermost entity is more likely to have its inner entities shared either the first token or the last token, making the constructed graph  more instructive, thus its inner entities are easier to extract.  
%Our model is also effective for long outermost entities, e.g., when the length of outermost entities is 15, the F1-score of inner entities is almost as high as that of their outermost entities.

\begin{table*}[!t] 
\centering
\begin{tabular}{lccccrccccr}
\toprule
\multirow{2}{*}{length}  && \multicolumn{4}{c} {outermost entities} && \multicolumn{4}{c} {{inner entities}} \\
 \cline{3-11}
&& {P} & {R} & {F}  & {Num.}  && {P} & {R} & {F}  & {Num.} \\
\hline
1 && 75.9    &80.6     & 78.2    & 1,260  && - & - & -&-\\
2 && 72.1    &74.8     & 73.4    &  488  && 76.6    &63.6     & 69.5 & 77 \\
3&& 67.8    &   72.2  &  69.9   &    198 && 67.6 & 56.5&  61.5 &85\\
4&&  62.5   &  60.9   &   61.7  &  112   && 68.1&  42.3&52.2 &111\\
5&&   60.7  &   48.7  &   54.0   &   76  && 56.0& 37.8& 45.1&74\\
6&&  46.3   &  46.3   &   46.3  &  41   &&  28.0& 25.9& 26.9&54\\
7&&   44.4  &  30.8   &   36.4  &  26    && 21.7& 16.7& 18.9&30\\
8&&   64.3  &   40.9  &   50.0  &    22 && 31.8& 21.2& 25.5&33 \\
9&&  35.7   &   31.3  &    33.3 &  16   && 23.1& 19.4& 21.1&31\\
10&& 57.1    &   22.2  &   32.0  &  18   && 20.0& 15.4& 17.4&26\\
\bottomrule
\end{tabular}
\caption{Length-wise results on ACE2005 test dataset.} 
\label{lengthwise}
\end{table*}

\subsection{Ablation Study}
In this paper, we use the interactions of flat module and graph module to respectively help better predict outermost and inner entities. We conduct ablation study to verify the effectiveness of the interactions.
The first part is the information delivery from the flat  module to the graph module.
We conduct four experiments: (1) no graph: we skip Eq. (5)-(6) and let graph feature $f = \Call{Linear}{x}$. In this case, inner entities are independent of the outermost entities and only rely on the word representation (section 2.1) which carries contextualized information. (2) adjacent graph: we further utilize the sequential information of the text to help inner entity prediction. (3) entity graph: the boundary information of outer entities can be indicative for inner entities, we construct an entity graph based on the entities extracted by the flat module. (4) both graphs: when outer entities are not recognized by the flat module, their inner entities will fail to receive the boundary information, we use the sequential information of the text to make up for the deficiency of using only entity graph. Experimental results show that entity graph carries more useful information than adjacent graph, which enhances the baseline by 1.4/1.1/1.2 $F_1$ score, respectively.
By combing these two graphs together, we get a larger gain of 1.7/1.6/1.6 $F_1$ score.
The second part is the information delivery from  the graph module  to  the flat module, the new representation $X^{new}$ learned from graph module is propagated back to the flat  module. $X^{new}$ is equipped with the dependencies of inner entities and shows useful, yielding an improvement of 0.8/1.5/0.5 $F_1$ for the three benchmarks, respectively.

\subsection{Inference Time}

We examine the inference speed of our BiFlaG with \cite{zheng2019boundary}, \cite{sohrab2018deep} and \cite{ju2018neural}  in terms of the number of words decoded per second. 
For all the compared models, we use the re-implemented code released by \cite{zheng2019boundary} and set the same batch size 10.
Compared with \cite{zheng2019boundary} and \cite{sohrab2018deep}, our BiFlaG
does not need to compute region representation for each potential entity, thus we can take full advantage of GPU parallelism.
Compared with \cite{ju2018neural}, which requires CRF decoding for each layer, our model only needs to calculate two modules, by contrast, the cascaded CRF layers limit their inference speed.

\begin{figure}[!t]
  \centering 
  \includegraphics[scale=0.55]{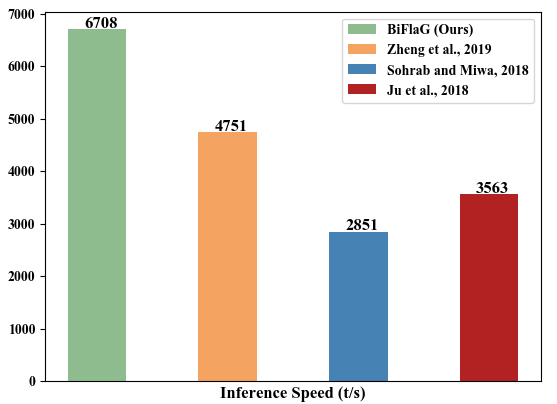}
    \caption{The inference speed of our BiFlaG and
compared models on GENIA test set. t/s indicates token per second.}\label{time}
\end{figure} 

\begin{table*}[thbp]
% \resizebox{1.0\columnwidth}{!}{
\centering
\begin{tabular}{|c|p{0.8\textwidth}|}  
\hline
Setence & Interesting aside: Starbucks is  taking over the location in my town that was recently abandoned by Krispy Kreme. \\
\hline
Gold Label & ORG: \{Starbucks, Krispy Kreme\}; FAC: \{the location in my town that was recently abandoned by Krispy Kreme; that\}; GPE: \{my town\}; PER: \{my\} \\
\hline
No Graph & ORG: \{Starbucks\}; LOC: \{the location in my town that was recently abandoned by Krispy Kreme\}; PER: \{my\} \\
\hline
No interaction & ORG: \{Starbucks, Krispy Kreme\}; LOC: \{the location in my town that was recently abandoned by Krispy Kreme\}; GPE: \{my town\}; PER: \{my\} \\
\hline
BiFlaG & ORG: \{Starbucks, Krispy Kreme \}; FAC: \{the location in my town that was recently abandoned by Krispy Kreme; that\}; GPE: \{my town\}; PER: \{my\} \\ 
\hline
\end{tabular}
\caption{An example of predicted results in ACE2005 test dataset.}
\label{case}
\end{table*} 

\section{Case Study}
Table \ref{case} shows a case study of each module in our model. In this example, entities \emph{my}, \emph{my town}, \emph{that} and \emph{Krispy Kreme} are nested in the entity \emph{the location in my town that was recently abandoned by Krispy Kreme}. Our BiFlaG model successfully extracts all these entities with exact boundaries and entity categorical labels. Without graph construction, nested entities \emph{my town}, \emph{that} and \emph{Krispy Kreme} are not identified. Without interaction between the two modules, the outermost entity \emph{the location in my town that was recently abandoned by Krispy Kreme} is mislabeled as LOC (location), which is actually a FAC (Facility) type, inner nested entities \emph{my}, \emph{my town} and \emph{Krispy Kreme} are not propagated back to the flat module, which maybe helpful to correct the extracting of the outermost entity.

\section{Related Work}

Recently, with the development of deep neural network in a wide range of NLP tasks \cite{bai-zhao-2018-deep,huang-etal-2018-moon,huang-zhao-2018-chinese,he-etal-2018-syntax,he-etal-2019-syntax,li-etal-2018-seq2seq,li-etal-2018-unified,li2019dependency,zhou-zhao-2019-head,xiao-etal-2019-lattice, zhang-zhao-2018-one,zhang-etal-2019-open,zhang2019dcmn+,zhang2020SemBERT,zhang2020sg}, it is possible to build reliable 
NER systems without hand-crafted  features.
Nested named entity recognition requires to identity all the entities in texts that may be nested with each other. 
Though NER is a traditional NLP task, it is not until the very recent years that researches have been paid to this nested structure for named entities.
 
\cite{lu2015joint} introduce a novel hypergraph representation to handle overlapping mentions.
\cite{muis2018labeling} further develop a gap-based tagging
schema that assigns tags to gaps between words to address the spurious structures issue, which can be modeled using conventional linear-chain CRFs. However, it suffers from the structural ambiguity issue during inference. 
\cite{wang2018neural} propose a novel segmental hypergraph
representation to eliminate structural ambiguity.
\cite{katiyar2018nested} also propose a hypergraph-based approach based on the BILOU tag scheme that utilizes an LSTM network to learn the hypergraph representation in a greedy manner.

Stacking sequence labeling models to extract entities from inner to outer (or outside-to-inside) can also handle such nested structures.
\cite{alex2007recognising} propose several different modeling techniques (layering and cascading) to combine multiple CRFs for nested NER. However, their approach cannot handle nested entities of the same entity type.
\cite{ju2018neural} dynamically stack flat NER layers, and recognize entities from innermost layer to outer ones. Their approach can deal with nested entities of the same type, but suffers from error propagation among layers. 

Region-based approaches are also commonly used for nested NER by extracting the subsequences in sentences and classifying their types.
\cite{sohrab2018deep} introduce a neural exhaustive model
that considers all possible spans and classify their types. This work is further improved by \cite{zheng2019boundary}, which first apply a single-layer sequence labeling model to identify the boundaries of potential entities using context information, and then classify these boundary-aware regions into their entity type or non-entity.
\cite{lin2019sequence} propose a sequence-to-nuggets
approach named as Anchor-Region Networks (ARNs) to detect nested entity mentions. They first use an anchor detector to detect the anchor words of entity mentions and then apply a region recognizer to identity the mention boundaries centering at each anchor word.
\cite{fisher2019merge} decompose nested NER into two
stages. Tokens are merged into entities through real-valued decisions, and then the entity embeddings are used to label the entities identified.

\section{Conclusion}
This paper proposes a new bipartite flat-graph (BiFlaG) model for nested NER which consists of two interacting subgraph modules. Applying the divide-and-conquer policy, the flat module is in charge of outermost entities, while the graph module focuses on inner entities. Our BiFlaG model also facilitates a full bidirectional interaction between the two modules, which let the nested NE structures jointly learned at most degree. As a general model, our BiFlaG model can also handle non-nested structures by simply removing the graph module. In terms of the same strict setting, empirical results show that our model generally outperforms previous state-of-the-art models.

\bibliography{acl2020}
\bibliographystyle{acl_natbib}

\end{document}